\documentclass[fleqn,usenatbib]{mnras}

\usepackage{newtxtext,newtxmath}

\usepackage[T1]{fontenc}

\DeclareRobustCommand{\VAN}[3]{#2}
\let\VANthebibliography\thebibliography
\def\thebibliography{\DeclareRobustCommand{\VAN}[3]{##3}\VANthebibliography}

\usepackage{graphicx}	
\usepackage{amsmath}	
\usepackage{dirtytalk}

\usepackage{lscape}
\usepackage{longtable}
\usepackage{booktabs}
\usepackage{ulem}
\usepackage{subfigure}
\usepackage{soul}

\usepackage{lineno}        

\usepackage{letltxmacro}
\LetLtxMacro{\oldtextsc}{\textsc}
\renewcommand{\textsc}[1]{\oldtextsc{\scalefont{1.3}#1}}

\usepackage{xargs}                      
\usepackage[pdftex,dvipsnames]{xcolor}  

\usepackage[colorinlistoftodos,prependcaption,textsize=tiny]{todonotes}
\newcommandx{\unsure}[2][1=]{\todo[linecolor=red,backgroundcolor=red!25,bordercolor=red,#1]{#2}}
\newcommandx{\change}[2][1=]{\todo[linecolor=blue,backgroundcolor=blue!25,bordercolor=blue,#1]{#2}}
\newcommandx{\info}[2][1=]{\todo[linecolor=OliveGreen,backgroundcolor=OliveGreen!25,bordercolor=OliveGreen,#1]{#2}}
\newcommandx{\improvement}[2][1=]{\todo[linecolor=Plum,backgroundcolor=Plum!25,bordercolor=Plum,#1]{#2}}
\newcommandx{\thiswillnotshow}[2][1=]{\todo[disable,#1]{#2}}

\title[The high state of EF Eri]{Phase-resolved spectroscopic observations of the magnetic cataclysmic binary EF Eridani: Revealing complex magnetic accretion during a high state\thanks{Based on observations made with the Southern African Large Telescope (SALT) under program 2021-2-LSP-001 (PI: D. A. H. Buckley)}}

\author[Khangale et al.]{
Z. N. Khangale,$^{1}$\thanks{E-mail: khangalezn@saao.ac.za (ZNK)}
S. B. Potter,$^{2,3}$
D. A. H. Buckley$^{1,2,4}$ and
P. E. Barrett,$^{5}$ 
\\
$^{1}$Department of Astronomy, University of Cape Town, Private Bag X3,
Rondebosch 7701, South Africa\\
$^{2}$South African Astronomical Observatory, Observatory Road, Observatory, 7925, Cape Town, RSA\\
$^{3}$Department of Physics, University of Johannesburg, PO Box 524, Auckland Park 2006, South Africa\\
$^{4} $Department of Physics, University of the Free State, PO Box 339, Bloemfontein 9300, South Africa\\
$^{5}$Department of Physics, The George Washington University, Washington, DC 20052 USA
}

\date{Accepted 2025 October 08. Received 2025 October 08; in original form 2025 June 24}

\pubyear{2025}

\begin{document}
\label{firstpage}
\pagerange{\pageref{firstpage}--\pageref{lastpage}}
\maketitle

\begin{abstract}
We present high-resolution, phase-resolved spectroscopic observations of the polar EF Eri, obtained with SALT and the SAAO 1.9-m telescope during its recent emergence from a three-decade-long low state. 
The average spectrum shows strong emission from the Balmer lines (H$\alpha$ and H$\beta$) and He~\textsc{ii} 4686 \AA{}, along with weaker emission from the He~\textsc{i} lines and the Bowen fluorescence (C~\textsc{iii}/N~\textsc{iii}) blend at 4650 \AA{}. 
The emission lines redward of 5500 \AA{} transition to pure absorption at orbital phases $\sim$0.75--0.95, which we attribute to obscuration of the line-emitting region by the accretion stream. 
Trailed spectra of the emission lines reveal multicomponent structures consistent with other polars. In this first Doppler study of EF Eri, tomograms of the strongest lines (He~\textsc{ii} 4686 \AA{} and the Balmer lines), using both the standard and inside-out projections, identify three key emission regions: the irradiated face of the secondary star, the ballistic and threading regions of the accretion stream, and the magnetically confined flow. Our Doppler maps show not only the ballistic stream but also two unambiguous magnetic accretion flows, which is consistent with the presence of multiple magnetic accretion regions.
\end{abstract}




\section{Introduction}

Magnetic cataclysmic variable (mCV) stars are mass-transferring binary systems consisting of a magnetic white dwarf (WD) primary accreting material from a late-type main-sequence secondary star, usually a red dwarf via a Roche lobe overflow \citep{1995cvs..book.....W}. 
mCVs comprise intermediate polars (IPs, sometimes also called DQ Her stars; \citealt{1956ApJ...123...68W, 1994PASP..106..209P}) and polars, or the so-called AM Her-type \citep[see e.g.][]{1977ApJ...211..859B,1977ApJ...212L.125T} systems. 
These systems are classified according to the magnetic field strength of their WDs and the degree of synchronisation of the binary and/or spin periods of the WDs. 
The field strength in IPs ranges between 4--30 MG \citep{2008ApJ...684..558P,2020AdSpR..66.1209D} and this is enough to disrupt and truncate the inner part of the accretion disk formed around the WD, or even prevent disk formation \citep{1995MNRAS.275.1028B}. 
In addition, the field strength in IPs is inadequate to force the two stars to synchronise in the binary period; instead, the WD spin period is less than or a fraction of the binary orbital period. 
It is worth mentioning that IPs may accrete via the accretion disk for disk-fed systems, directly along magnetic field lines for disk-less systems, or in a hybrid mode in the form of disk overflow \citep[see][and references therein]{2020AdSpR..66.1209D}. 
On the other hand, polars have higher field strengths, ranging between 10--230 MG \citep[see e.g.][]{2020AdSpR..66.1025F,2024A&A...684A.190B} and this prohibits the formation of an accretion disk as well as synchronising the two stars rotational and orbital periods. 
The binary's orbital period and the WD spin period are the same in this scenario. 
However, there are more than twelve polars in which the WD spin is slightly out of synchronism, i.e. the two periods differ by less than 2\% (e.g. V1432 Aql \citep{1995MNRAS.273..681W,1996A&A...306..860F}, BY Cam \citep{1997MNRAS.290...25S,1998MNRAS.295..511M}) or where the desynchronisation differs by more than 3\% (e.g. IGR J19552$+$0044 \citep{2017A&A...608A..36T} and Paloma \citep{2007A&A...473..511S}). 
Polars exhibit low and high states of accretion, and the transition from high to low states is quite sudden because they do not have a readily available supply of material stored in an accretion disk. 
Therefore, these stars offer us an opportunity to study the dynamics of accretion and how strong magnetic fields influence the flow of matter.

EF Eri is a well-studied non-eclipsing mCV, first identified as a polar by \cite{1979ApJ...232L..27G} and \cite{1979Natur.281...48W}, with its classification subsequently confirmed through detections in optical of  circular polarisation reaching $+$9\% \citep{1979IAUC.3324....2H, 1979IAUC.3327....2H}. 
This binary system has an orbital period of 81 minutes \citep[see e.g.][]{1980Natur.285..306B,1979IAUC.3324....2H}.   
It has been studied extensively, through either spectroscopy or photometry and/or polarimetry, in different wavelengths, including optical \citep[e.g.][]{1979ApJ...234L.113B, 1981MNRAS.196..425B}, infrared \citep[e.g.][]{1979IAUC.3335....1W,1981MNRAS.195..155A, 2004ApJ...614..947H,2006ApJ...646L..65H}, ultraviolet \citep[e.g.][]{1981Natur.290..119C, 2006ApJ...646L.147S, 2010ApJ...716.1531S}, X-rays \citep[e.g.][]{1981ApJ...244L..85W, 1987ApJ...316..360B, 2007A&A...469.1027S} and radio \citep{2020AdSpR..66.1226B}. 
EF Eri, like other polars, shows short-term and long-term brightness variations. 
The short-term variations include rapid flickering with amplitudes of 0.2 magnitude, flaring events with amplitudes $\sim$0.8 magnitude \citep[see e.g.][]{1979IAUC.3335....1W, 1979ApJ...234L.113B} and sometimes 6-minute optical and infrared quasi-periodic oscillations (QPOs) as first reported by \cite{1980PASP...92..178W}. 
The high-speed photometric study conducted by \cite{1982ApJ...252..277W} shows the absence of 6-minute QPOs; however, additional non-periodic oscillations with periods of 13.7 and 20 minutes were present.  
Similar short-term variations, including flaring events \cite[e.g.][]{1981ApJ...244L..85W} and 6-minute QPOs \citep{1987ApJ...316..360B}, were also reported in the soft and/or hard X-ray light curves of EF Eri. 
However, QPOs were not seen in the X-ray light curves presented by \cite{1991A&A...246L..36B}, but these authors reported aperiodic soft X-ray flaring events. 
Recent hard X-ray observations in the high state, presented by \cite{2025ApJ...987...53F}, show no evidence of QPOs. 

The long-term variations arise from either an increase or a decrease in mass transfer from the donor star, resulting in the high and low states of accretion we observe. 
During the high state, the magnitudes in the $V$ band increase to $\sim$14 magnitude \citep{1982MNRAS.199..801B} while during the low states, the magnitudes of the $V$ band drop to lower than 18 \citep[see e.g.][]{1998ASPC..137..446W, 2006ApJ...652..709H}. 
EF Eri has always been among the X-ray bright CVs \citep[e.g.][]{1981ApJ...244L..85W,1981ApJ...245..618P} and was not seen in a low state from its discovery in the late 1970s until the late 1990s. 
Before the current high state, EF Eri spent an extended period in a low state, which spanned $\sim$25 years, starting in 1997 \citep[see e.g.][]{1998ASPC..137..446W} and lasting up until December 2022. 
This low state was preceded by a long high state, which lasted over 18 years (1979–1997). 
There are two reported rebrightening episodes, such as those reported by \cite{2006ApJ...652..709H} and \citep{2008ATel.1562....1M}, the latter lasting $\sim$2 months. 
Note that flaring events and/or frequent rebrightening episodes have been observed in other polars, e.g. AM Her and AR UMa \citep{2024ApJ...965...96M}. 
Detailed multi-band photometric and polarimetric studies of EF Eri obtained during the low, intermediate, and high states have been presented in the literature \citep[see e.g.][references therein]{1979ApJ...234L.113B, 1982MNRAS.199..801B, 1985MNRAS.212..709C, 1987A&A...186..120P, 2003AJ....125.2609H, 2006ApJ...646L.147S, 2010ApJ...716.1531S}.  

The first optical spectra of EF Eri were presented by \cite{1979ApJ...232L..27G} and \cite{1979Natur.281...48W}, and their spectra revealed strong emission lines from the Balmer series as well as emission from neutral and singly-ionised Helium lines. 
The presence of strong Balmer and Helium emission lines in polars signifies a high state of accretion. 
Further optical spectroscopic observations of EF Eri obtained during the high state were presented by different authors \citep[see e.g.][]{1980A&A....86L..10V, 1980ApJ...238..946S, 1981MNRAS.196..425B, 1985MNRAS.212..609M}. 
Spectroscopic results from these studies revealed the variability of emission lines with the 81-minute orbital period; multi-component and sinusoidal radial velocity curves with high-velocity amplitudes of $\sim$800--1200\,km\,s$^{-1}$; complex emission line profiles -- consisting of a narrow peak superposed on a broad base -- through contour maps as well as multiple components within the emission lines. 
The complex emission profiles were attributed to emission from two emitting regions between the two stars, separated by less than 180 degrees. 
In addition, \cite{1985MNRAS.212..609M} established that the main source of emission lines in EF Eri was likely the accreting gas and attributed the source of the narrow component as likely due to X-ray heating of either the secondary or material around the inner Lagrangian ($L_1$) point. 

The low-state optical spectra of EF Eri were initially presented by  \cite{1998ASPC..137..446W} and show weak emission lines from the Balmer lines superposed on a steep blue continuum -- with Zeeman absorption features around the H$\alpha$ and H$\beta$ lines and these are reminiscent of the photosphere of the WD. 
Further analysis of the low-state optical spectra performed by \cite{2000A&A...354L..49B} revealed no trace of emission from secondary star in the red part of the spectrum implying a later-type spectral type than M9 for the companion star. 
The mid-IR observations by Spitzer of EF Eri \citep{2006ApJ...646L..65H} established that the secondary star is an L or T brown dwarf.
Follow-up low-state optical spectra of EF Eri were presented by different authors between 2000 and 2010 \citep[see e.g.][]{2002ASPC..261...57H, 2003AJ....125.2609H, 2004ApJ...614..947H, 2006ApJ...652..709H} and shows significant absorption around the H$\beta$ line and prominent Zeeman features. 
The field strengths determined from Zeeman absorption features is $\sim$13 MG \citep[e.g.][]{1998ASPC..137..446W,2007A&A...463..647B}, whereas field strength determined from modelling of cyclotron features gives 16.5 and 21 MG for the two cyclotron emitting regions \citep{1996A&A...306..860F}.
The observations presented by \cite{2003AJ....125.2609H} showed weak emission from the H$\beta$ line and this was lower than the continuum -- implying that this system was in an even lower state than recorded by \cite{1998ASPC..137..446W}.   
In addition, the optical spectra obtained between 2002--2003 and presented by \cite{2004ApJ...614..947H} show a steep blue continuum with absorption lines and no Balmer emission lines were visible. 
The Zeeman absorption components $\sigma^{\pm}$ of the Balmer lines were visible in the spectra. 
In contrast, the optical spectra obtained between 2004 August and 2006 February and presented by \cite{2006ApJ...652..709H} show the presence or return of the Balmer series and HeI lines in emission with underlying Zeeman-splitting absorption lines from the photosphere of the WD. 
However, no He~\textsc{ii} emission lines were observed in this system's spectra. 
Similar low-state optical spectra of EF Eri were presented by \cite{2010ApJ...716.1531S} and \cite{2010A&A...514A..89S} and showed weak Balmer emission superposed on broad absorption lines. 
The distance to EF Eri from, as derived from Gaia DR3  \citep{2016A&A...595A...1G, 2023A&A...674A..38G} parallax measurements, is approximately 160 $\pm$ 4 pc, which, despite uncertainties, aligns with the \cite{2003AJ....126.3017T} estimate and the upper limit established by \cite{2010A&A...514A..89S}.

In this paper, we present detailed high-time resolution spectroscopic observations of EF Eri obtained during the recent high state and use phase-resolved spectra to construct Doppler tomograms. Specifically, we present both the standard Doppler maps and flux modulation Doppler tomography techniques. 
This article is outlined as follows: Section \ref{section:observations} describes the observations and data reduction carried out as well as details of the instruments used. 
The results are described in detail in Section \ref{section:results_and_analysis} starting with the spectral analysis in Section \ref{section:spectrocopy} and
Section \ref{section:Doppler_Tomography} presents all the Doppler maps produced and finally, Section \ref{section:discussion} contains the discussion and summary of the results.

\section{Observations}\label{section:observations}

\begin{table*}
    \caption{Spectroscopic log of observations for EF Eri.}
    \label{tab:observations}
    \centering
    \begin{tabular}{c c c c c c c c c c } \\ \hline
    Date of     & Telescope  & Instrument & Type of    & Grating & Grating & Wavelength &Exposure & Number & Phase \\
    observation &  used      & used       & observation&  used   & angle ($^{\circ})$  & range (\AA{}) &  time (s)  & of spectra  & coverage \\ \hline \hline
        2023-01-26       & SALT       & RSS    & spectroscopy      & PG1300 
        & 21.5 & 4550--6630 & 30 & 100 & 0.186--0.804 \\
        2023-01-27       & SALT       & RSS    & spectroscopy      & PG1300 & 21.5 & 4550--6630 & 30 & 100  & 0.903--1.514 \\
        2023-02-01        & SAAO 1.9-m & SPUPNIC & spectroscopy    &  Grating 4 & 3.6 & 4200--5500 & 200 & 56 & 0.921--3.415 \\
        2023-02-02        & SAAO 1.9-m & SPUPNIC & spectroscopy    &  Grating 4 & 3.6 & 4200--5500 & 200 & 54 & 0.935--3.237 \\
        2023-02-04        & SAAO 1.9-m & SPUPNIC & spectroscopy    &  Grating 4 & 3.6 & 4200--5500 & 200 & 54 & 0.252--2.611 \\ \hline
    \end{tabular}
\end{table*}

\subsection{Spectroscopy with SALT}\label{section:observations_salt}

We obtained spectroscopic observations of EF Eri with the Southern African Large Telescope (SALT; \citealt{2006SPIE.6269E..0AB}) using the Robert Stobie Spectrograph (RSS; \citealt{2003SPIE.4841.1463B,2003SPIE.4841.1634K}) on the night of 2023 January 26 and 27, respectively. The RSS instrument was used in frame-transfer mode. 
The observations cover $\sim$0.61 of the orbital phase of EF Eri per night but due to unforeseen events, the phase bins between 0.8 and 0.9 were not covered.
The PG1300 grating was used and the grating angle was set at $\sim$21.5$^{\circ}$ to give a wavelength range from $\approx$4550 to 6630 \AA{}. 
This setup ensured sufficient coverage of the He~\textsc{ii} 4686 \AA{} and H$\alpha$ emission lines.  
Exposure times of 30 s ($\triangle \phi$ = 0.0061) were used for each of the science frames improving on earlier observations of this target which used exposure times of $\geq$200 s \cite[e.g.][]{1980ApJ...238..946S}.
Wavelength calibration utilised CuAr arc exposures made at end of the science exposures.  
Observations of the spectrophotometric standard star, EG21, was taken with the same setup as the science exposures on the night of 2023 January 27 for flux calibration purposes. 

The CCD pre-processing of the observations was performed using the \textsc{pysalt}\footnote{For more details on pysalt visit http://pysalt.salt.ac.za/.} package \citep{2010SPIE.7737E..25C}, this includes overscan correction, bias subtraction, gain correction as well as cosmic ray removals. 
The remainder of the reductions such as wavelength calibration and spectral extraction were carried out using \textsc{iraf}\footnote{IRAF is distributed by the National Optical Astronomy Observatories, which are operated by the Association of Universities for Research in Astronomy, Inc., under a cooperative agreement with the NSF. The IRAF astronomical data reduction and analysis software package is available at \url{https://iraf-community.github.io}.} and/or \textsc{pyraf}. 
This includes performing optimal extraction \citep{1986PASP...98..609H} of the science and standard star spectra with subtraction of the sky background. 
Relative flux correction of the science exposures was based on the sensitivity of the spectrophotometric standard star EG21. All the science spectra were corrected for atmospheric extinction. 
Figure \ref{figure:average-spectrum} shows the average flux-calibrated spectra of EF Eri obtained from the two nights. This will be discussed in Section \ref{section:spectrocopy}.  
The two gaps in the spectra, of width $\sim$35 \AA{}, are due to the RSS detector consisting of a mosaic of three chips.

\subsection{SPUPNIC spectroscopy}\label{section:observations_spupnic}

Further observations of EF Eri were obtained with the South African Astronomical Observatory (SAAO) 1.9-m telescope using the Spectrograph Upgrade: Newly Improved Cassegrain (SPUPNIC; \citealt{2016SPIE.9908E..27C, 2019JATIS...5b4007C}) instrument during the three nights between 2023 February 1 and 2023 February 4, respectively. We used grating number 4 and a grating angle of 3.6, giving a wavelength coverage of $\sim$4200 \AA{} to 5500 \AA{}. A slit width of 1.8${''}$ was used, and the exposure time was set to 200 s ($\triangle \phi$ = 0.041) to achieve a good signal-to-noise ratio. 
Each observations cover over two cycles of the orbital phase of EF Eri per night.
Spectra of the CuAr arc, with an exposure time of 10s, were taken at the start of the observations and then at frequent intervals (i.e. after eight or nine science exposures) using the same grating as the science exposures for wavelength calibration purposes. Spectra of spectrophotometric standard stars were obtained at the end of each night with the same grating for flux calibration. The data were reduced using \textsc{iraf} and/or \textsc{pyraf} routines and \textsc{python} scripts incorporating \textsc{Astropy} routines. These included overscan subtraction, flat-fielding, and bias correction. The one-dimensional spectra were extracted using the \textsc{apall} task in \textsc{iraf}.

\section{Results and analysis}\label{section:results_and_analysis}

\begin{figure}
    \centering
    \includegraphics[width = 0.45\textwidth]{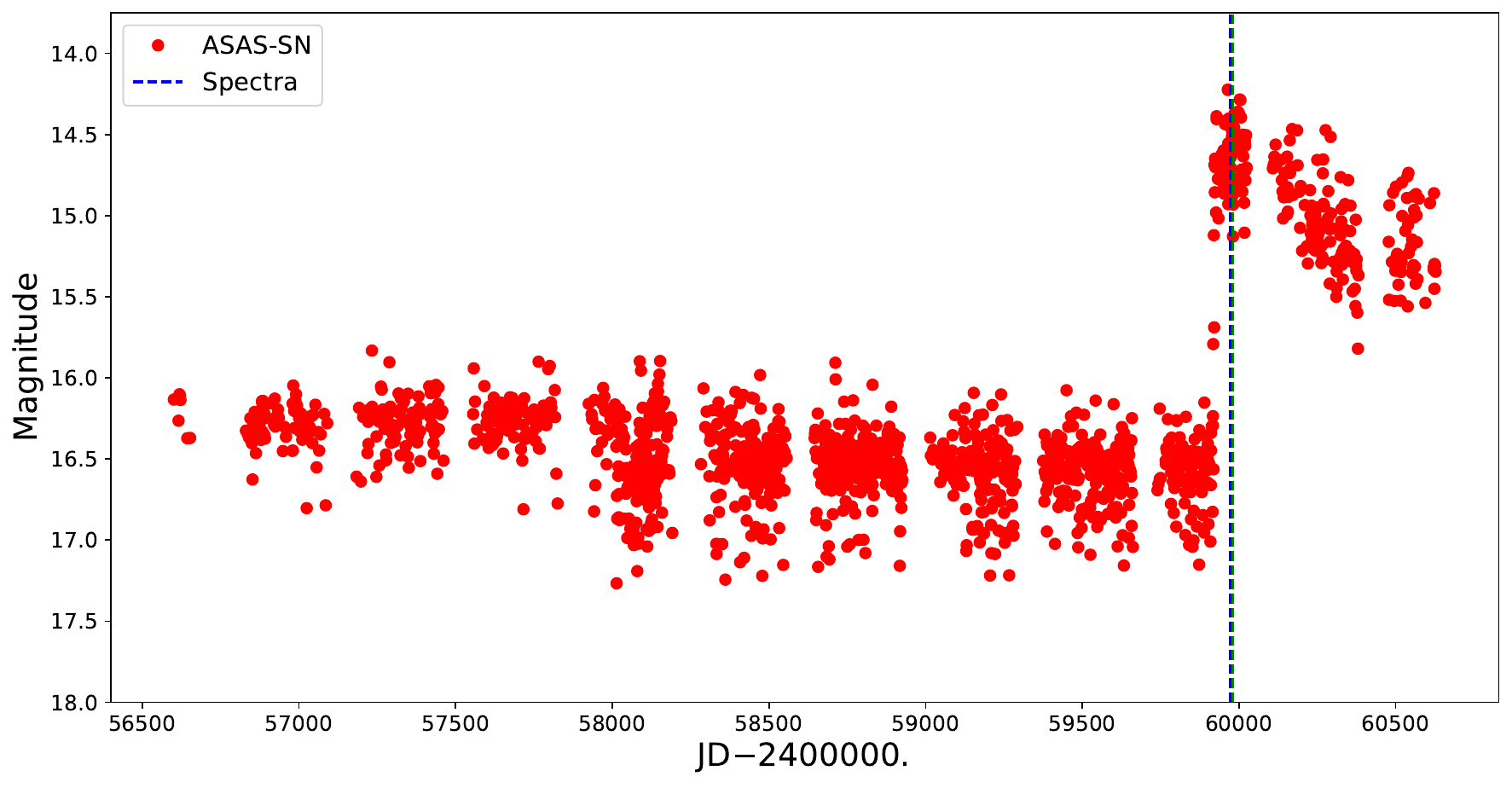}
    \caption{Long-term ASAS-SN light curve of EF Eri spanning 11 years between 2013 November 5 and 2025 January 30 showing the extended low-state and the recent high-state. The vertical blue and green dashed lines mark the dates of the spectroscopic observations with SALT and the SAAO 1.9-m telescope, respectively.}
    \label{figure:long-term-lc}
\end{figure}

Figure~\ref{figure:long-term-lc} shows the long-term photometric light curve of EF Eri retrieved from the All-Sky Automated Survey for SuperNovae (ASAS-SN; \citealt{2014ApJ...788...48S,2017PASP..129j4502K,2023arXiv230403791H}) spanning ten years, from 2013 November 4 to 2025 January 30. It is evident from the figure that EF Eri has spent most of the past decade in a low state, with magnitudes ranging between $\sim$17.5 and 15.5. However, a recent excursion to a high state was observed on 2022 December 7--9, during which the magnitude of EF Eri increased by about 1.5 magnitudes to $\sim$14.5 mag, and it has remained in a high state since then.

Note that the spectroscopic observations presented in this paper were obtained during the peak of this high state, which is marked with a blue dashed line in Figure~\ref{figure:long-term-lc}. The most recent photometric observations, obtained around January 2025, indicate that the source is now gradually declining from the high state. Continued photometric monitoring of this target is therefore essential to determine how long EF~Eri remains in the high state, and whether a rebrightening event might occur before the system returns to a low state.

\subsection{Spectroscopic ephemeris}

The ephemeris of EF~Eri has been the subject of several papers in the literature and has been determined in various ways: based on the linear polarisation pulse \citep{1979IAUC.3327....2H}; the ephemeris derived by \cite{1981MNRAS.196..425B} (and refined by \citealt{1982MNRAS.199..801B}) using ten sharp infrared minima measured from photometric light curves \citep{1981MNRAS.195..155A,1980Natur.285..306B}; and also based on the maximum light in photometric light curves \citep{1980MNRAS.191P..43W}. Given the different methods and the lack of a consistent ephemeris, it is difficult to compare or match different datasets. For example, the ephemeris of \cite{1981MNRAS.196..425B} required an addition of 0.14 and 0.42 in phase to convert their phases to spectroscopic and magnetic phases, respectively. In another example, \cite{1982MNRAS.199..801B} suggested that their infrared ephemeris could be converted to polarimetric phases by adding 0.43 to the resulting orbital phases calculated using their ephemeris. 

Other methods used to derive the epoch of EF~Eri include radial velocity fitting, adopting the time of the blue-to-red crossing as phase zero. For instance, \cite{1985MNRAS.212..609M} used the zero-crossing of the Balmer line radial velocities to derive an epoch from their velocity curves, which they then combined with literature timings to produce a more accurate ephemeris. \cite{2010A&A...514A..89S} presented the latest ephemeris of EF~Eri, based on the blue-to-red radial velocity zero crossing of the Ca \textsc{ii} 8542 \AA{} emission line from observations obtained in 2009 August. However, the limited precision of the period combined with the long time span makes this ephemeris unusable for our new observations.

\begin{figure}
\begin{center}$
\begin{array}{cc}
\includegraphics[height = 11cm ]{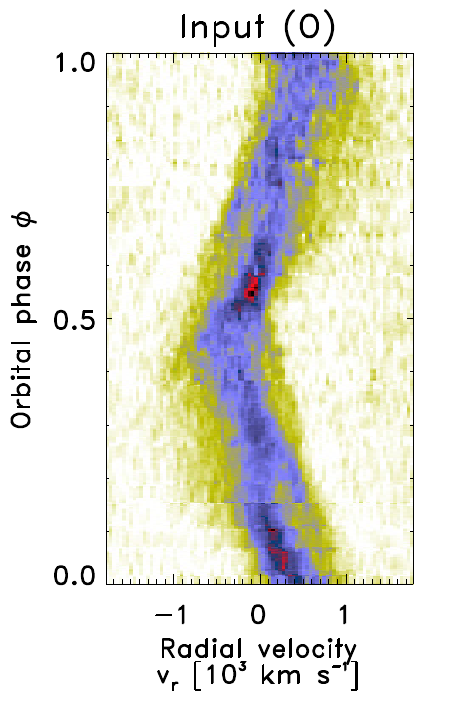}
 \hspace{-0.75cm}
\raisebox{1.0cm}{\includegraphics[ height= 9.6cm]{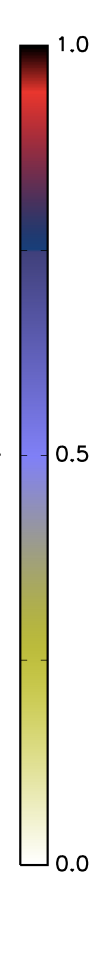}} \\ 
\end{array}$
\end{center}
\caption{The observed trailed spectra of EF Eri based on the He~\textsc{ii} 4686 \AA{} line from the observations taken with the SAAO 1.9-m telescope. The colour bar to the right shows the scale with which the trailed spectra were produced. }
\label{figure:ephemeris}
\end{figure}

We used the observed trailed spectra of the He~\textsc{ii}~4686~\AA{} line from the SAAO 1.9-m telescope observations to derive a new epoch which was then used to phase all our observations. Figure~\ref{figure:ephemeris} shows the observed trailed spectrum of He~\textsc{ii} 4686 \AA{}. 
The trailed spectra have been normalised by the maximum flux in the input spectra, and the colour bar to the right shows the scale with which the trailed spectra were produced. 
Multiple components are visible, which will be discussed in Section~\ref{section:trailed_spectra}. We identified a faint and narrow, redshifted, low-velocity emission component between phases 0.3 and 0.5, respectively. 
The narrow component was first identified in the optical spectra of polars by \cite{1991A&A...244..373S, 1993A&A...267..103S} for MR Ser and V834 Cen and later reported in the high-state spectra of HU Aqr \citep{1995RvMA....8..125S}. 
It is understood to originate from emission lines emitted from the face of the secondary star that is illuminated by ultraviolet and soft X-rays from the WD. This component is moving from red to blue and crosses zero velocity at phase 0.5. Therefore, phase 0 would then correspond to the inferior conjunction of the secondary, from which we derive a new orbital ephemeris:
\begin{equation}
{\rm{BJD(TDB)}} = 2459977.2250(30) + 0.05626586(8)E,
\end{equation}
where $E$ is the cycle number and BJD is the barycentric Julian date in the barycentric dynamical time (TDB) system. 
We adopted the orbital period and its uncertainty from \cite{1982MNRAS.199..801B}.  
The uncertainty in the epoch ($T_0$) was estimated from the width of the orbital phase bins, which represents the temporal resolution of the variability, and was subsequently converted into an uncertainty in $T_0$. 
This ephemeris and orbital period were used to assign phases to all remaining spectroscopic observations. All spectroscopic times were first converted from Julian dates to BJD using the prescriptions of \cite{2010PASP..122..935E}.

\subsection{Spectroscopy}\label{section:spectrocopy}

\begin{figure*}
    \centering
    \includegraphics[width = 0.95\textwidth ]{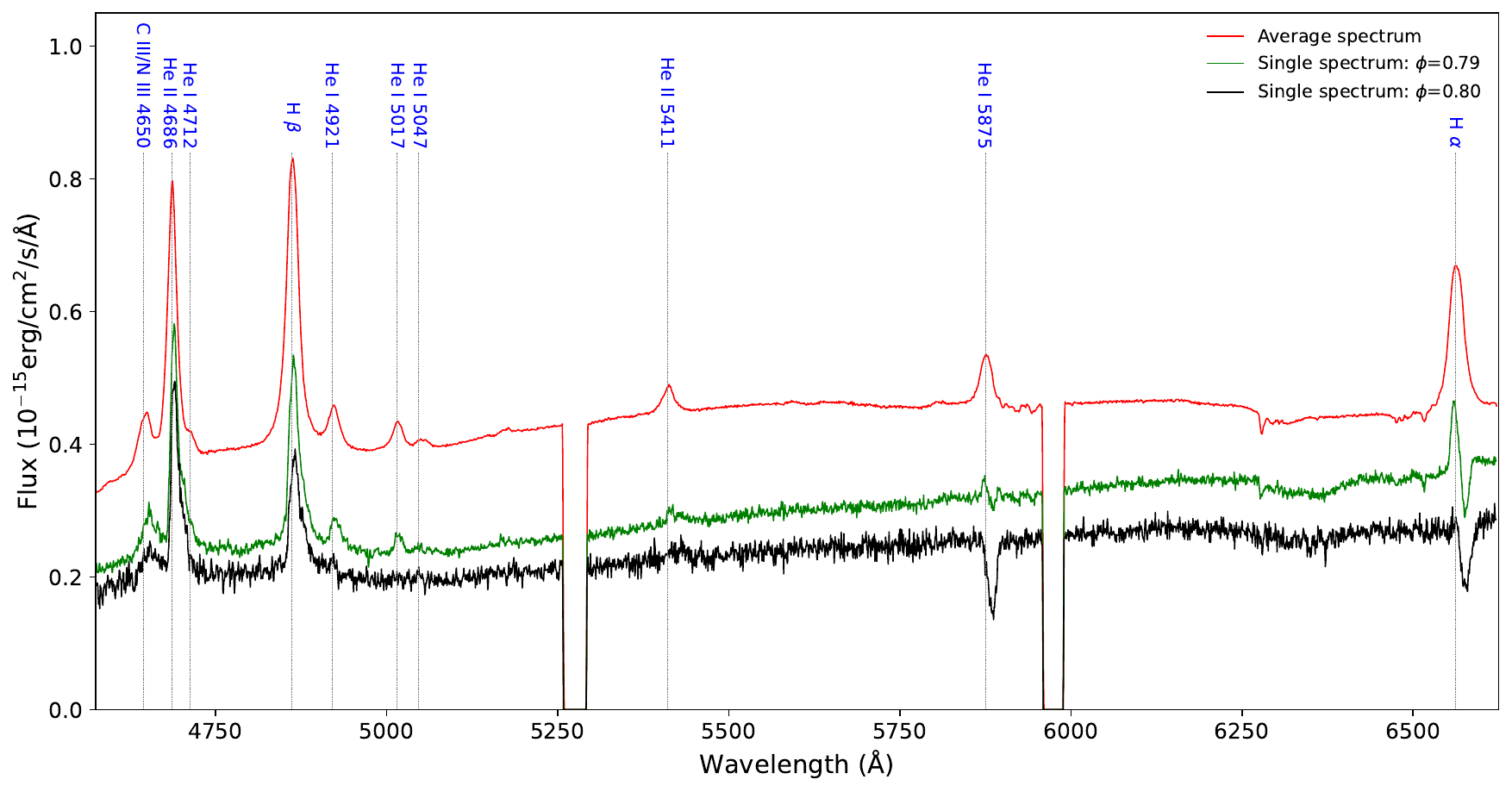}
    \caption{Average flux-calibrated SALT spectrum (red line) of EF Eri obtained on the nights of 2023 January 26 and 27, respectively. 
    The spectra in green and black show the individual exposures corresponding to phase 0.79 and 0.8 and highlight the reversals in emission lines. 
    Prominent emission lines have been labelled.}
    \label{figure:average-spectrum}
\end{figure*}

The high state spectra of EF Eri has been discussed in detail in literature \citep[see e.g.][]{1979ApJ...232L..27G,1979Natur.281...48W,1980A&A....86L..10V} and  shows strong emission from the Balmer and He~\textsc{ii} 4686 \AA{} lines.
Figure \ref{figure:average-spectrum} shows the average spectrum of EF Eri (solid red line) obtained using SALT from the observations taken on 2023 January 26 and 27, respectively. 
The spectrum displays characteristic emission lines of AM Her-type stars in a high state, with strong, broad single-peaked Balmer  (H$\alpha$ and H$\beta$) and He~\textsc{ii} 4686 \AA{} emission lines. 
Other emission lines present include He~\textsc{i} (at 5875 \AA{}, 5017 \AA{}, 5047 \AA{} and 4921 \AA{}), He~\textsc{ii} at 5411 \AA{} and the Bowen blend (C~\textsc{iii}/N~\textsc{iii}) at 4650 \AA{} within the observed wavelength range.  
All emission lines showed variations with the orbital phase, notably the H$\alpha$ and He~\textsc{i} 5875 \AA{} lines switch from emission to absorption (see the solid green and black lines) lines from phase $\sim$0.73 to $\sim$0.98, respectively. 
This phenomenon, previously reported by \cite{1980A&A....86L..10V}, is attributed to the eclipse of the line-emitting region by the accretion curtain. 
Other alternative explanations for emission line reversals include obscuration by the threading region or absorption in the secondary star's atmosphere. The former is ruled out since EF Eri is a non-eclipsing system and the threading region can not cross our line of sight since this region confined to the orbital plane \citep[see][for further discussion]{2018AJ....155...18L}. 
The absorption by the secondary star's atmosphere is also not a plausible explanation as such absorption would occur at phase 0.0. 
EF Eri is one of seven polars that show \say{pre-eclipse} emission line reversals in their optical spectra; others include MN Hya \citep{1998MNRAS.301...95R}, FL Cet \citep{2005ApJ...620..422S}, V808 Aur \citep{2016AstBu..71..101B}, MASTER OT J132104.04$+$560957.8 \citep{2018AJ....155...18L}, ZTF J0850$+$0443 \citep{2023ApJ...945..141R} and 1RXS J184542.4$+$483134 \citep{2023AstL...49..706K}. 
Typically, these reversals are observed in the Balmer and He\,\textsc{i} lines. However, exceptions include V808 Aur, where the He~\textsc{ii}~4686 \AA{} line exhibits an absorption core, and MASTER~OT~J132104.04+560957.8, which shows red wing absorption in the same line. 
Notably, the redshifted component goes to absorption first, except in FL Cet, where the blue wing is absorbed first \citep[see e.g.][]{2005ApJ...620..422S}.
It is worth mentioning that five of these seven polars (MN Hya, FL Cet, V808 Aur, 1RXS J184542.4$+$483134 and ZTF J0850$+$0443) are eclipsing, suggesting that emission line reversals are more common in high-inclination systems. 
The strength or intensity of the He~\textsc{ii} 4686 \AA{} line is comparable to that of the H$\beta$ line signifying that EF Eri was observed in the high state. 
It is worth mentioning that we also observed variability in the continuum of EF Eri, i.e. the continuum changes shape from flat to steep in the red part of the spectrum, consistent with previous studies from literature \citep[see e.g.][]{1980ApJ...238..946S}.
Other polars, such as Gaia23cer \citep{2024AstL...50..335K}, also exhibit a steep continuum in the red part of its spectra.

\subsection{Doppler tomography}\label{section:Doppler_Tomography}

\begin{figure*}
\begin{center}$
\begin{array}{ccc}
\includegraphics[height= 8cm, width=8cm]{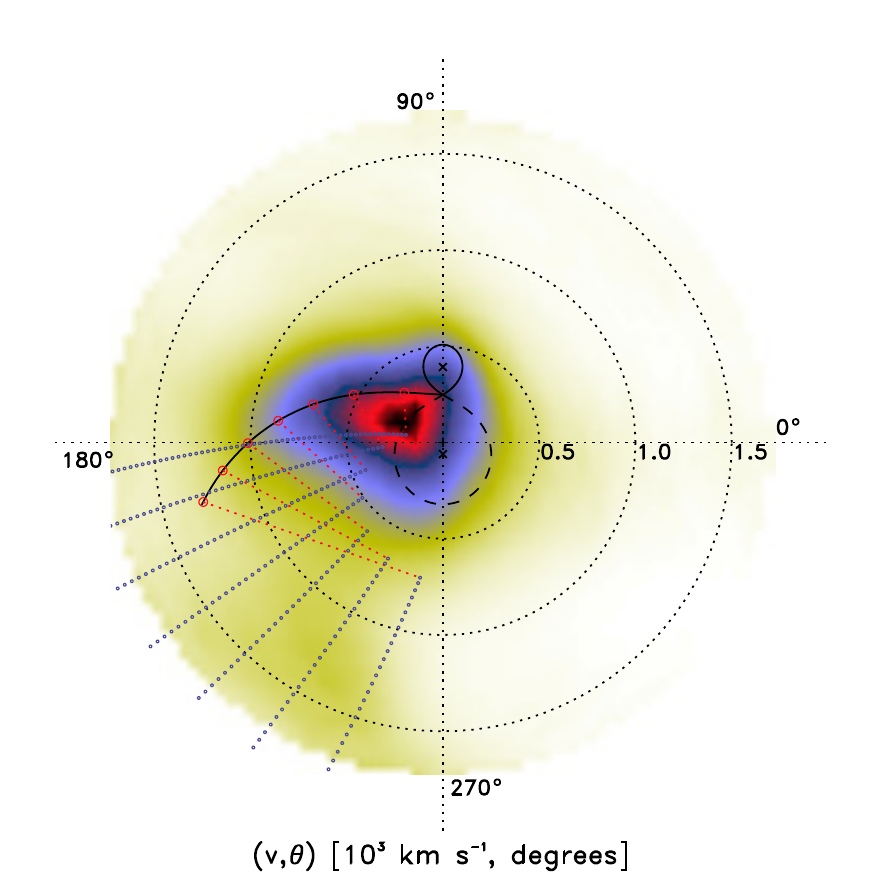}
 \hspace{-0.25cm}
\includegraphics[height= 8cm, width=8cm]{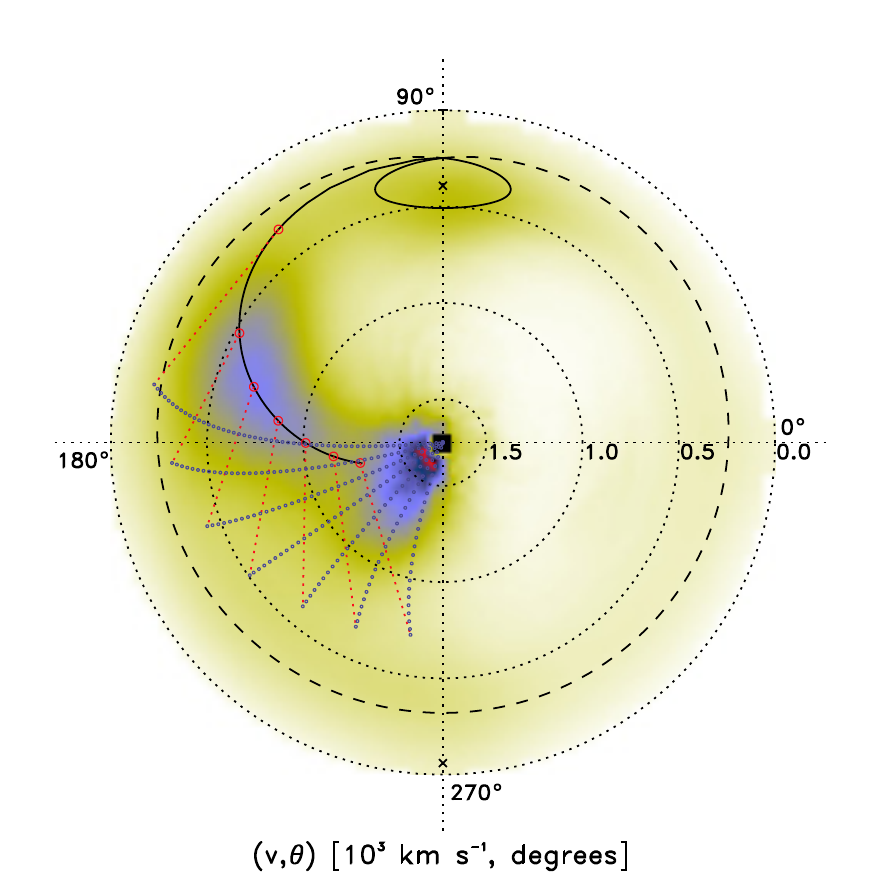}
 \hspace{-0.15cm}
\raisebox{-0.1cm}{\includegraphics[ height= 8.cm]{Spectra_colorbar_final.png}} \\
\end{array}$
\end{center}
\vspace{-0.15cm}
\begin{center}$
\begin{array}{cccc}
\includegraphics[width=0.3\textwidth, height= 8.5cm]{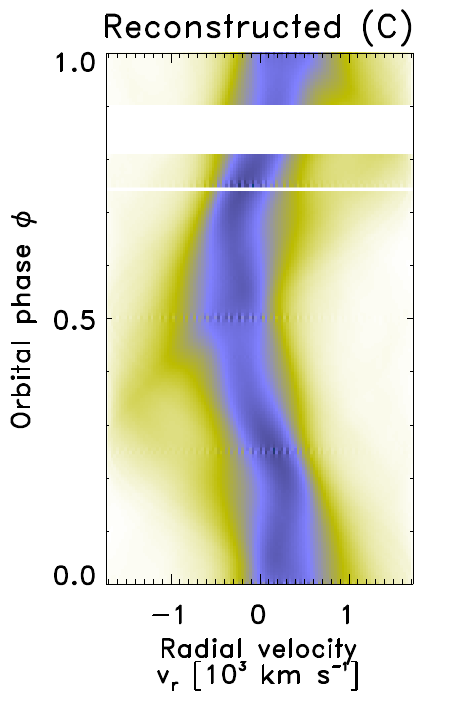} 
 \hspace{-.10cm}
\includegraphics[width=0.3\textwidth, height= 8.5cm]{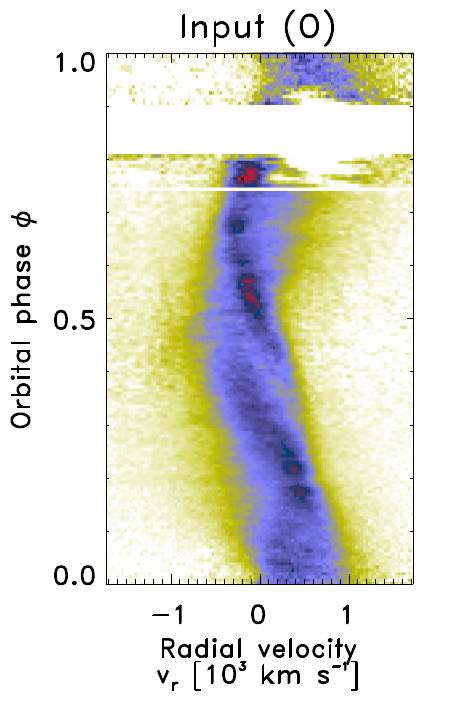} 
 \hspace{-.10cm}
\includegraphics[width=0.3\textwidth, height= 8.5cm]{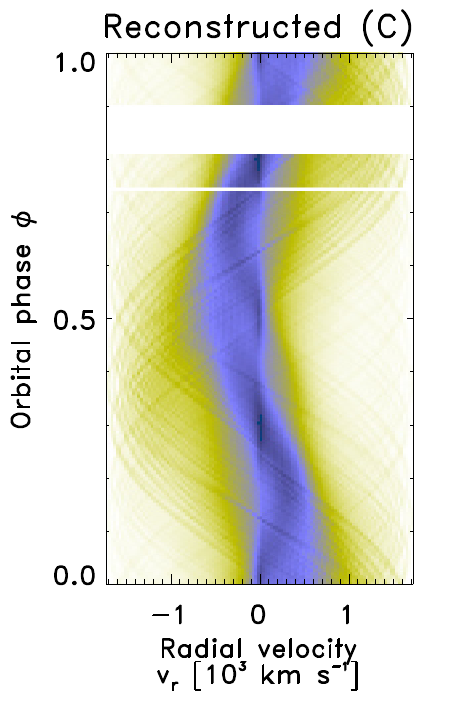}
 \hspace{-0.15cm}
\raisebox{0.75cm}{\includegraphics[ height= 7.5cm]{Spectra_colorbar_final.png}} \\ 
\end{array}$
\end{center}
\caption{Standard and inside-out Doppler tomograms as well as the observed and reconstructed trailed spectra based on the H$\alpha$ emission line from SALT \textbf{observations}. Top row: the standard (left) and inside-out (right) Doppler tomograms. Bottom row: observed trailed spectra (centre) and the corresponding reconstructed trailed spectra based on the standard (left) and inside-out (right) tomograms.}
\label{figure:DopplerHalpha}
\end{figure*}

\begin{figure*}
\begin{center}$
\begin{array}{ccc}
\includegraphics[height= 8cm, width=8cm]{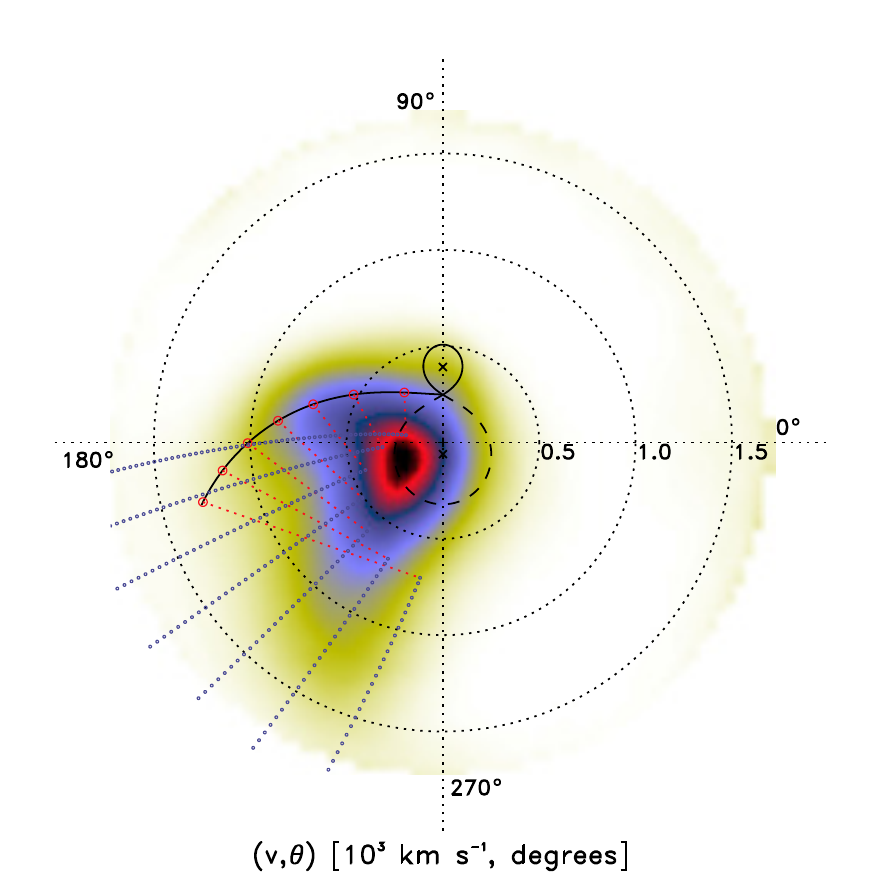}
 \hspace{-0.25cm}
\includegraphics[height= 8cm, width=8cm]{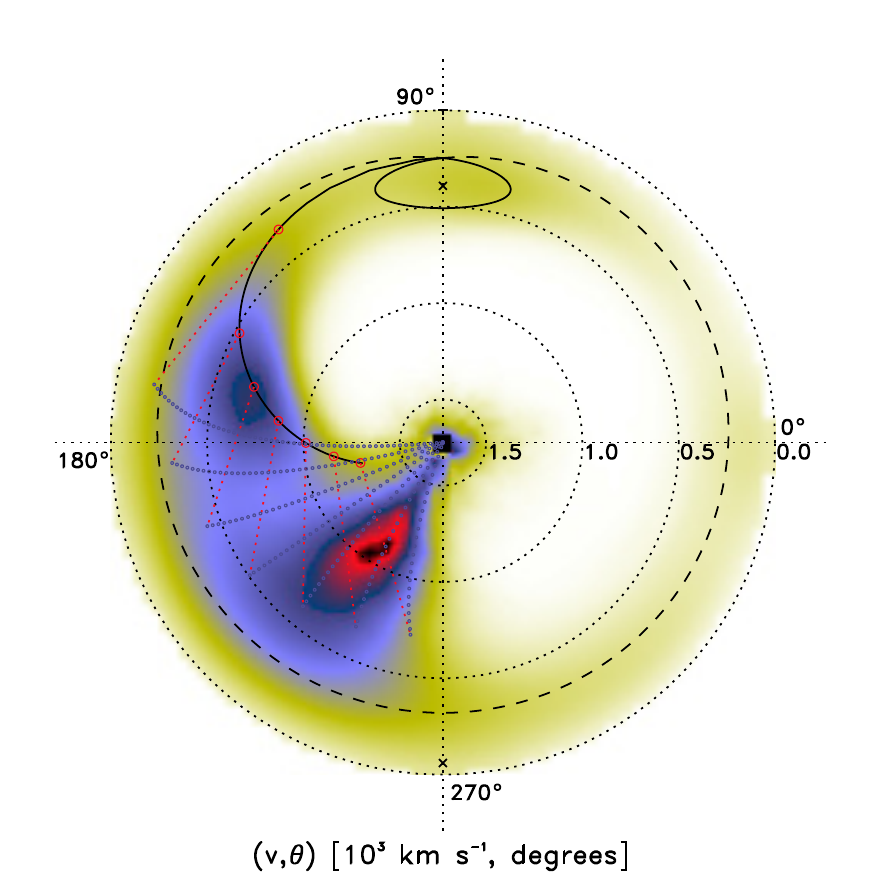}
 \hspace{-0.15cm}
\raisebox{-0.10cm}{\includegraphics[ height= 8.cm]{Spectra_colorbar_final.png}} \\ 
\end{array}$
\end{center}
\vspace{-0.15cm}
\begin{center}$
\begin{array}{cccc}
\includegraphics[width=0.3\textwidth, height= 8.5cm]{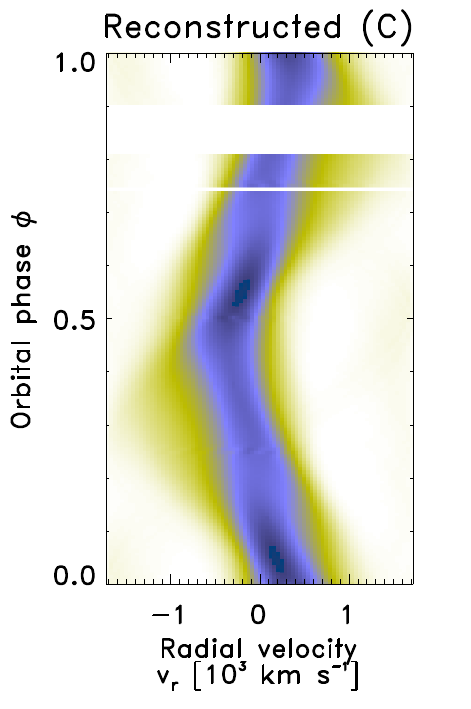} 
 \hspace{-.10cm}
\includegraphics[width=0.3\textwidth, height= 8.5cm]{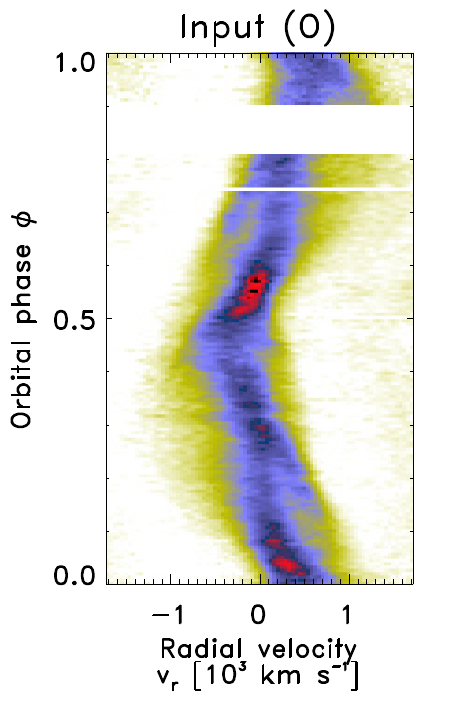} 
 \hspace{-.10cm}
\includegraphics[width=0.3\textwidth, height= 8.5cm]{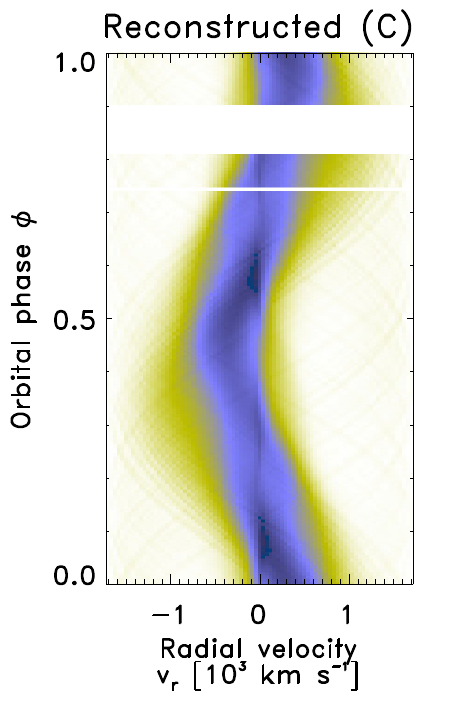}
 \hspace{-0.15cm}
\raisebox{0.75cm}{\includegraphics[ height= 7.5cm]{Spectra_colorbar_final.png}} \\ 
\end{array}$
\end{center}
\caption{Standard and inside-out Doppler tomograms as well as observed and reconstructed trailed spectra based on the He~\textsc{ii} 4686 \AA{} emission line from SALT observations. Top row: the standard (left) and inside-out (right) Doppler tomograms. Bottom row: observed trailed spectra (centre) and the corresponding  reconstructed trailed spectra based on the standard (left) and inside-out (right) tomograms.} 
\label{figure:DopplerHeII4686_New}
\end{figure*}

We used the strongest emission features from the high state spectra of EF Eri to compute Doppler maps using the \textsc{doptomog} code \citep{2015A&A...579A..77K,2016A&A...595A..47K}, which implements the maximum entropy regularization technique. 
In addition, the \textsc{doptomog} also incorporates the flux modulation mapping introduced by \cite{2003MNRAS.344..448S}, which extracts further information by utilizing the emission line flux variability over the orbital period. 
We analysed the Doppler maps corresponding to the strongest features: H$\alpha$, He~\textsc{i} 5875 \AA{}, He~\textsc{ii} 5411 \AA{}, H$\beta$ and He~\textsc{ii} 4686 \AA{}, but here we present only the Doppler maps of the H$\alpha$ and He~\textsc{ii} 4686 \AA{} emission lines from SALT observations. 
The top row of Figures \ref{figure:DopplerHalpha} and \ref{figure:DopplerHeII4686_New} show the standard (left) and inside-out projection (right) tomograms for the H$\alpha$ and He~\textsc{ii} 4686 \AA{} emission lines, respectively. 
The bottom row of the same figures show the observed trailed spectra (centre), standard reconstructed trailed spectra (left) and the inside-out reconstructed trailed spectra (right) for the same emission lines. 
To aid with the interpretation of the Doppler maps, we have overlaid a model with the WD mass, $M_1$ = 0.85 $M_{\odot}$, the mass ratio ($q = \frac{M_2}{M_1}$) of 0.154 and inclination, $i$ = 55$^{\circ}$ \citep[e.g.][]{1987Ap&SS.130..197P}. 
In the standard projection Doppler maps, the centre of mass of the binary system is indicated by a plus ($+$) sign and serves as the centre of the map. 
In the inside-out projection Doppler maps, the centre of mass of the binary system is the zero-velocity outer circumference of the tomogram. 
The centre of mass of both the primary and secondary are marked with a cross ($\times$) in all the Doppler maps.  
The Roche lobe of the WD is delineated with a dashed line,  while the secondary Roche lobe is depicted with a solid line in both the standard and inside-out Doppler tomograms. 
The ballistic stream trajectory is indicated by a solid line extending from the $L_1$ point up to 50 degrees in azimuth around the primary. 
The magnetic dipole trajectories are represented by thin dotted blue lines calculated at 10-degree intervals in azimuth. 
The colour-bars to the right of both the tomograms and trailed spectra, show the scale with which the Doppler maps and trailed spectra were produced.
We start by discussing the observed and reconstructed trailed spectra of the H$\alpha$ and He~\textsc{ii} 4686 \AA{} emission lines in Section \ref{section:trailed_spectra}.


\subsubsection{Observed and reconstructed trailed spectra}\label{section:trailed_spectra}

The bottom row of Figure \ref{figure:DopplerHalpha} shows the observed trailed spectra (centre) of the H$\alpha$ line, which show a relatively simple structure with at least two distinct sinusoidal components moving with a phase lag of $\sim$0.2 relative to one another.
The first is a broad component with mid-velocity amplitude, which is visible from phase $\approx$0.1--0.5 and crosses the zero-velocity from red to blue at phase $\sim$0.3.  
It is associated with emission from the ballistic stream, including the vicinity of the threading region.  
The second is a narrow component with low-velocity amplitude, which is visible from phase 0.4--0.9 and crosses the zero-velocity at phase $\sim$0.5. 
This component is associated with emission from the irradiated face of the secondary star \citep[see e.g.][]{1995RvMA....8..125S, 1995ASPC...85..166S}. 
The third is a relatively broad feature (yellow colour) visible throughout the orbital phase and has a maximum blueshift at phase 0.4--0.5 and a maximum redshift at phase 0.85. 
The two broad components (blue and yellow) are associated with emissions originating from various parts of the accretion flow, including the magnetically confined accretion stream \citep[see][]{1995RvMA....8..125S, 1995ASPC...85..166S}. 
Similarly, the bottom row of Figure \ref{figure:DopplerHeII4686_New} shows the observed trailed spectra (centre) of He~\textsc{ii} 4686 \AA{} emission line, which show a complex structure with at least three distinct emission components. 
But in this case, the low- and mid-velocity components (blue and red) move with a phase lag of $\sim$0.15 relative to each other. 
These two components blend together into a single broad component at some orbital phases except phases $\sim$0.15--0.3 and $\sim$0.65--0.8, respectively. 
These components cross the zero-velocity from red to blue at phase $\sim$0.3 and cross the zero-velocity from blue to red at phase $\sim$0.8. 
We associate these components with emissions from various parts of the accretion flow, including the ballistic stream (vicinity of the threading region) and the magnetically confined accretion stream \citep[see e.g.][]{1995RvMA....8..125S, 1999MNRAS.310..189F}. 
Similar to the Balmer lines, the third is a relatively broad feature (yellow colour), which is visible throughout the orbital phase. 
The broad base component is associated with emissions produced from various parts of the accretion flow, including the irradiated face of the secondary star. 
The bottom left and right panels of Figures \ref{figure:DopplerHalpha} and \ref{figure:DopplerHeII4686_New} show the standard and inside-out reconstructed trailed spectra of the H$\alpha$ and He~\textsc{ii} 4686 \AA{} lines, respectively. 
The reconstructed trailed spectra of both the H$\alpha$ and He~\textsc{ii} 4686 \AA{} emission lines reproduce the basic structure of the observed trailed spectra.

\subsubsection{Standard and inside-out projection tomograms}\label{section:doppler_maps}

The standard Doppler maps of the H$\alpha$ and He~\textsc{ii} 4686 \AA{} emission lines (top left panels of Figures~\ref{figure:DopplerHalpha} and \ref{figure:DopplerHeII4686_New}) are dominated by a \say{blob-like} feature of blended emission concentrated in the top and bottom left regions of the tomograms. 
The overall shape of the blended emission is different from one line to another. For the H$\alpha$ line, the blob-like feature is centred in the top left region of the tomogram and is associated mostly with unresolved emission from the ballistic stream and the irradiated face of the secondary star. There is evidence of some emission from the magnetically confined accretion stream in the bottom left region of the tomogram.
In contrast, the blob-like emission from the He~\textsc{ii} 4686 \AA{} line is centred in bottom left region of the tomogram and is associated with blended emission from the ballistic and magnetically confined accretion streams. The latter is indicated by the fanning diagonal lines in the lower left region of the tomogram and extending to higher velocities. 
The Roche lobe of the secondary star is overlaid in yellow colour for He~\textsc{ii} 4686 \AA{} and suggests that some emission from the secondary star is seen, although this is blended with the more dominant emission from the ballistic stream. 
Similar Doppler maps of other polars showing blended emission have been presented in literature e.g. BL Hyi, SRGA~J213151.5$+$491400 and ZTF~J0112$+$5827   \citep{1999ASPC..157...71S, 2024A&A...684A.190B, 2025A&A...694A.112L}. 

The inside-out projection tomograms (top right panels of Figures~\ref{figure:DopplerHalpha} and \ref{figure:DopplerHeII4686_New}) of the H$\alpha$ and He~\textsc{ii} 4686 \AA{} emission lines show more clearly separated emission features. For both lines, there is low-level emission around the expected position of the secondary star, located near the top centre of the tomograms.
The solid black line starting from the $L_1$ point traces the trajectory of the ballistic stream from the secondary star, which brightens upon reaching the threading region indicated by the red points. It is evident from the inside-out tomograms that the threading region is more pronounced in the case of the He~\textsc{ii} 4686 \AA{} line. At this region, the accretion stream slows down and is deflected to move perpendicular to the binary’s motion, following the magnetic field lines. This motion is indicated by the red dotted lines in the inside-out tomograms—magnetic dipole trajectories calculated at increments of 10$^{\circ}$ from 5$^{\circ}$ to 65$^{\circ}$ in azimuth around the WD—which trace the path of the magnetic stream as it leaves the orbital plane of the binary. The small blue circles mark the magnetic dipole trajectory as the stream converges to higher velocities while falling toward the WD.
The inside-out tomograms of both the H$\alpha$ and He~\textsc{ii} 4686 \AA{} lines show strong evidence of emission from various parts of the magnetically confined accretion stream in the lower left region of the tomograms. As already noted, the magnetically confined accretion stream dominates the emission in the He~\textsc{ii} 4686 \AA{} line, supported by the presence of a curtain-like feature spanning a wide range of velocities, from $\sim$250\,km\,s$^{-1}$ to $\geq$1500\,km\,s$^{-1}$. For the H$\alpha$ line, only the high-velocity part of the magnetically confined stream (toward the centre of the tomogram) is visible, and no distinct curtain-like structure is observed. 
The various components contained in the Doppler maps of polars (including EF Eri) were first identified by \cite{1997A&A...319..894S} in the tomograms of HU Aqr during high state. 
The Doppler maps discussed so far represent the phase-averaged emission distribution from the binary system. As a result, any phase-dependent variations, such as flux modulation, are not recovered in the reconstructed trailed spectra. These effects are explored in Section~\ref{section:amplitude_modulation} below, given the excellent quality of the data obtained for this target.


\subsubsection{Standard and inside-out modulation amplitude maps}\label{section:amplitude_modulation}

\begin{figure*}
\begin{center}$
\begin{array}{ccc}
\includegraphics[height= 7cm, width=7cm]{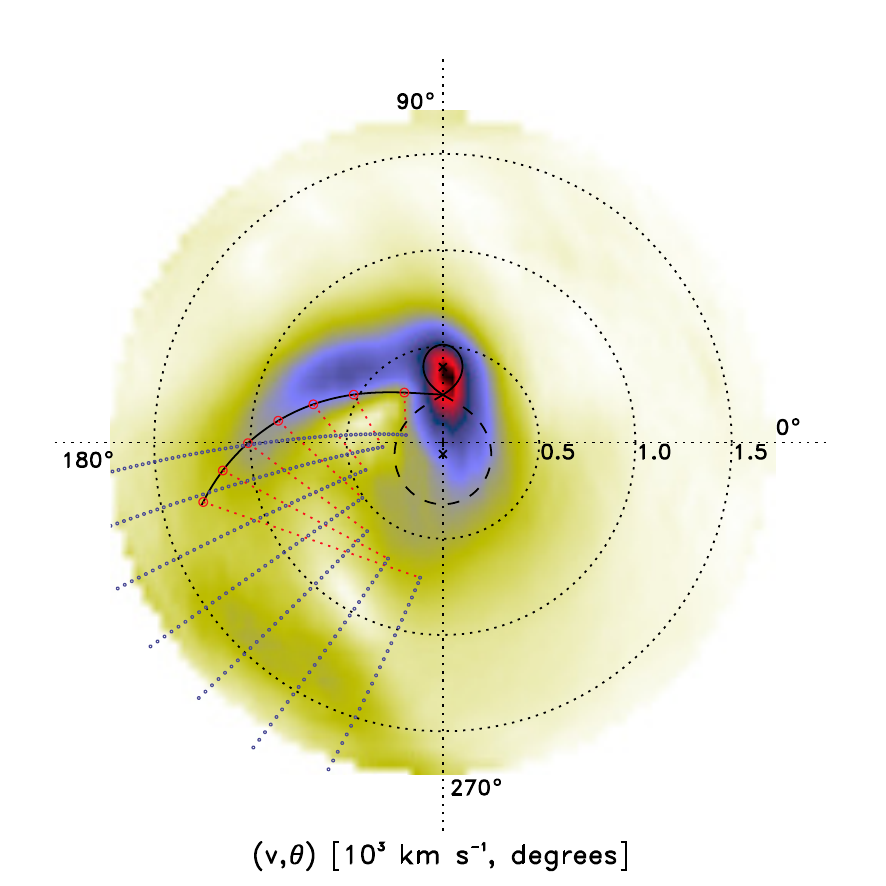} 
 \hspace{-0.25cm}
\includegraphics[height= 7cm, width=7cm]{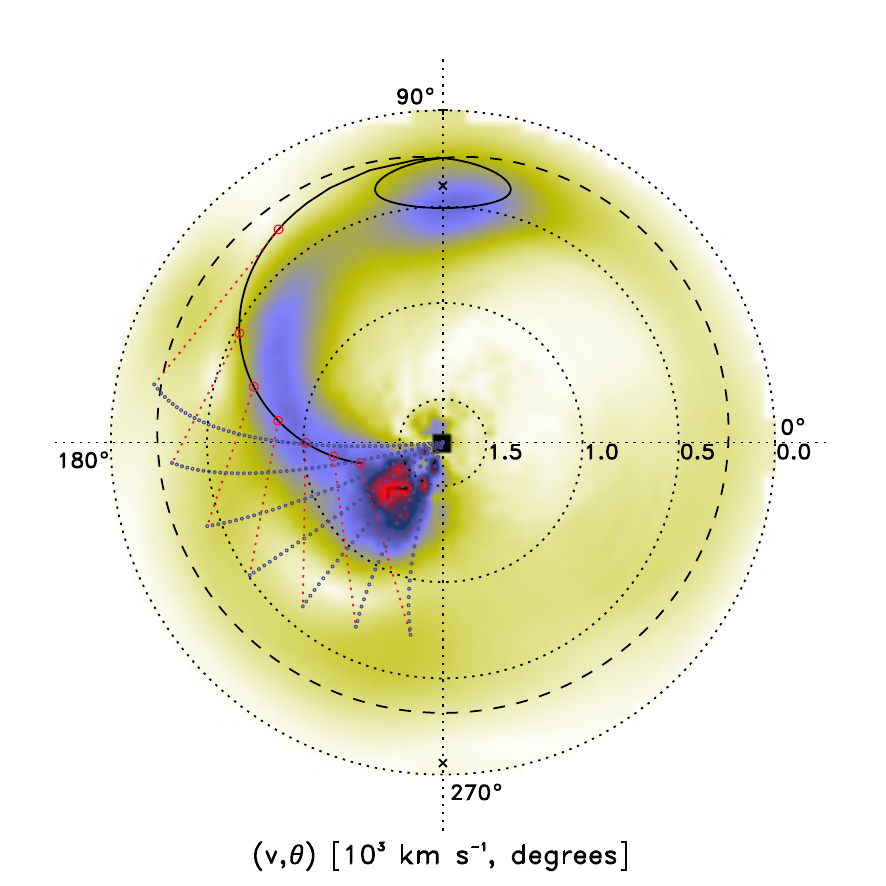}
 \hspace{-0.15cm}
\raisebox{-0.30cm}{\includegraphics[ height= 7.cm]{Spectra_colorbar_final.png}} \\ 
\end{array}$
\end{center}
\vspace{-0.35cm}
\begin{center}$
\begin{array}{ccc}
\includegraphics[height= 7cm, width=7cm]{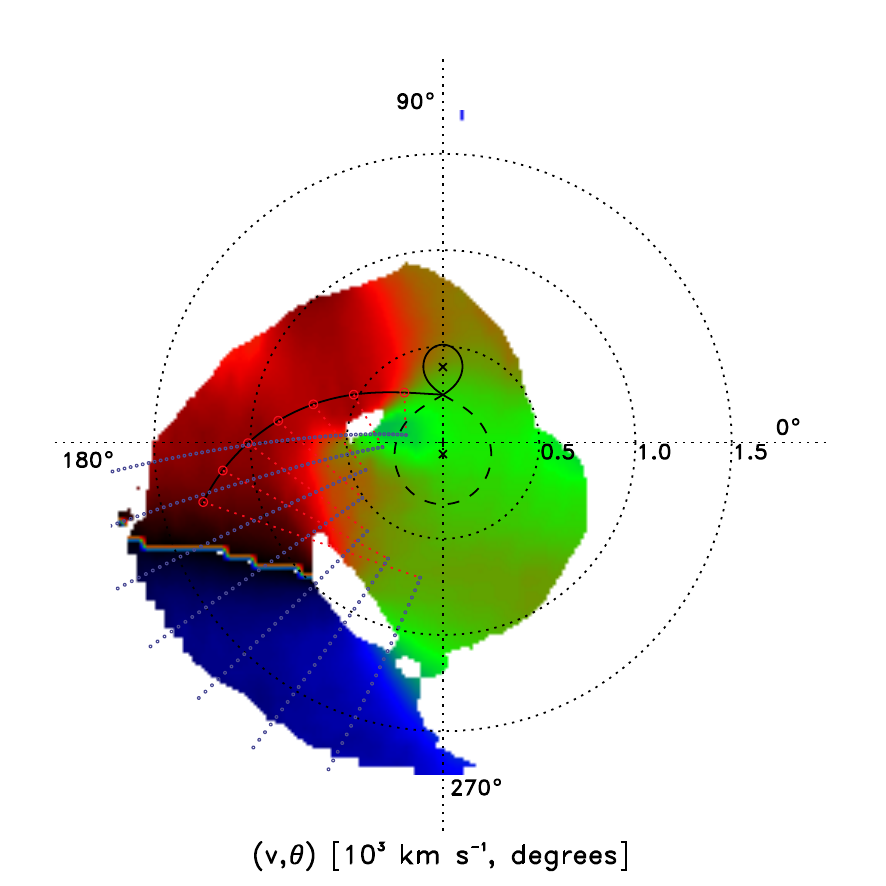} 
 \hspace{-0.250cm}
\includegraphics[height= 7cm, width=7cm]{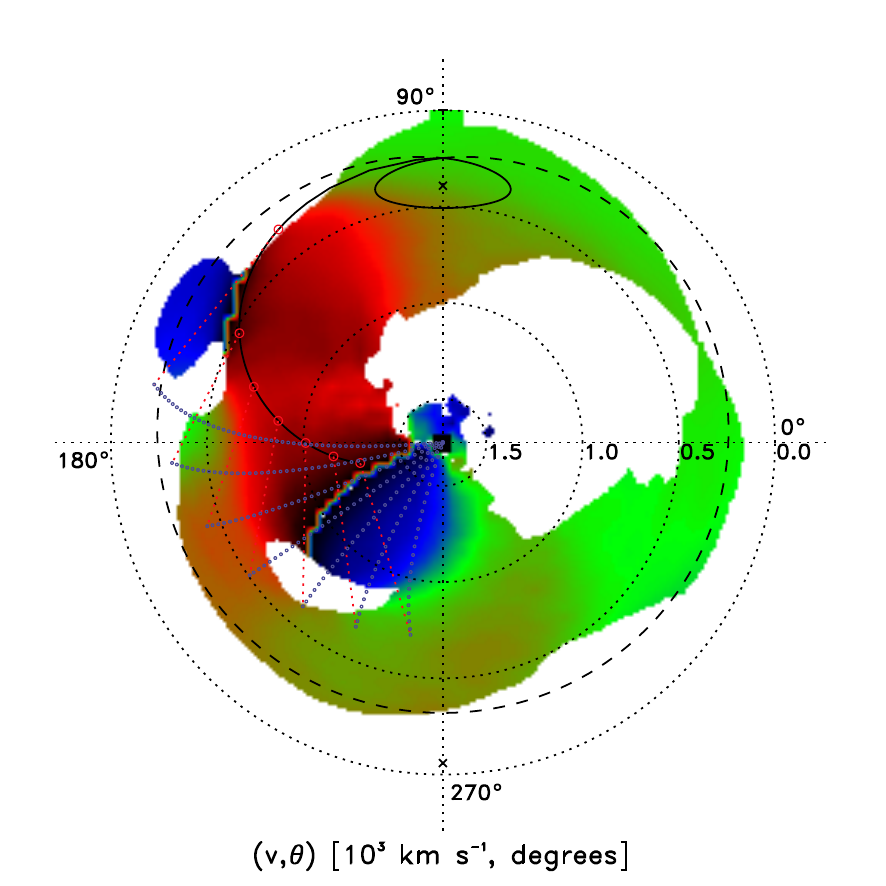}  
\hspace{-0.150cm}
\raisebox{-0.10cm}{\includegraphics[height= 7cm ]{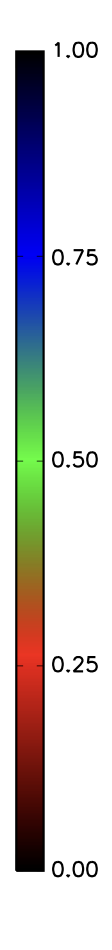}}  \\
\end{array}$
\end{center}
\vspace{-0.15cm}
\begin{center}$
\begin{array}{cccc}
\includegraphics[width=0.26\textwidth, height= 7.cm]{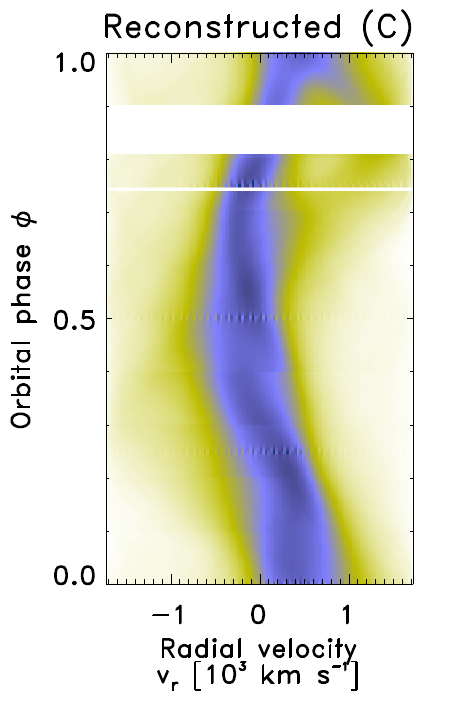} 
 \hspace{-.10cm}
\includegraphics[width=0.26\textwidth, height= 7.cm]{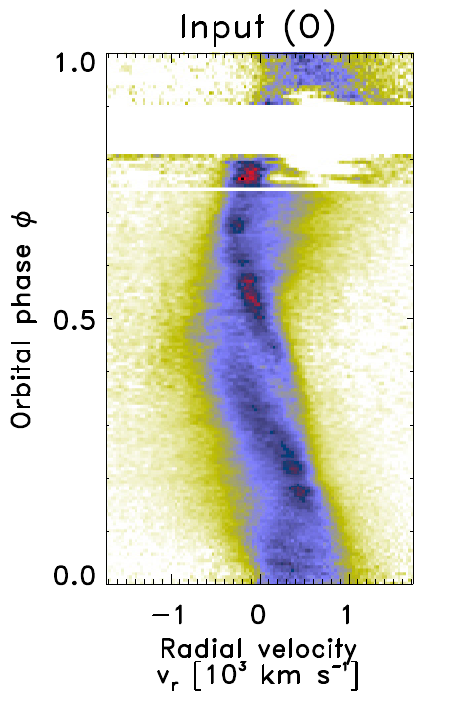} 
 \hspace{-.10cm}
\includegraphics[width=0.26\textwidth, height= 7.cm]{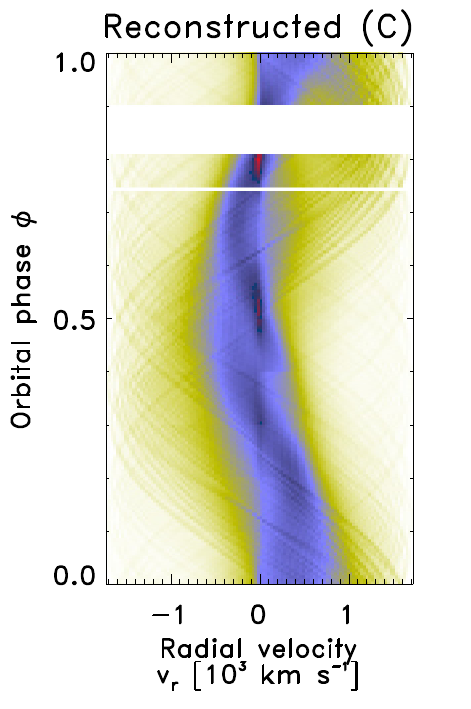} 
 \hspace{-0.15cm}
\raisebox{0.6cm}{\includegraphics[ height= 6.2cm]{Spectra_colorbar_final.png}} \\ 
\end{array}$
\end{center}
\vspace{-0.15cm}
\caption{Standard and inside-out Doppler maps together with the observed and reconstructed trailed spectra based on the H$\alpha$ emission line from SALT observations. Top row: the standard and inside-out flux modulation amplitude Doppler maps. Middle row: the standard and inside-out phase of maximum flux Doppler maps. The phase of maximum flux map shows only pixels where the corresponding modulation amplitude is at least 10\% of the maximum amplitude. Bottom row: the input trailed spectra (centre) with the summed reconstructed trailed spectra for the ten consecutive half-phases for standard (left) and inside-out (right), respectively.}
\label{figure:Halpha_mod}
\end{figure*}

\begin{figure*}
\begin{center}$
\begin{array}{ccc}
\includegraphics[height= 7cm, width=7cm]{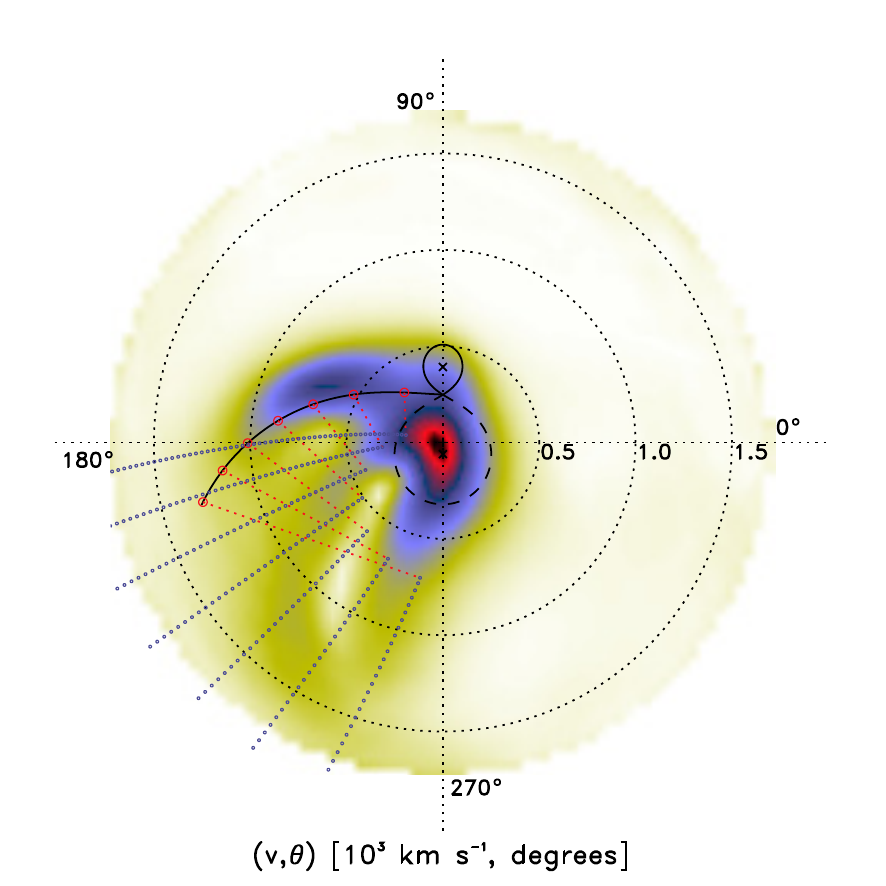} 
 \hspace{-0.25cm}
\includegraphics[height= 7cm, width=7cm]{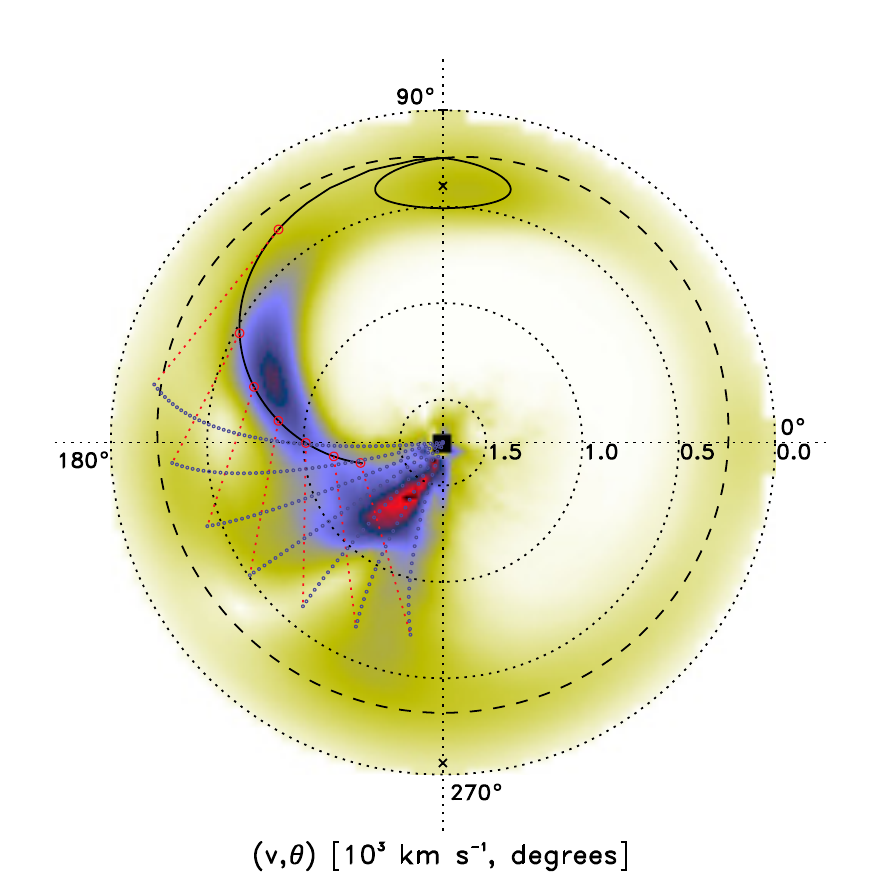}
 \hspace{-0.15cm}
\raisebox{-0.10cm}{\includegraphics[ height= 7.cm]{Spectra_colorbar_final.png}} \\ 
\end{array}$
\end{center}
\vspace{-0.35cm}
\begin{center}$
\begin{array}{ccc}
\includegraphics[height= 7cm, width=7cm]{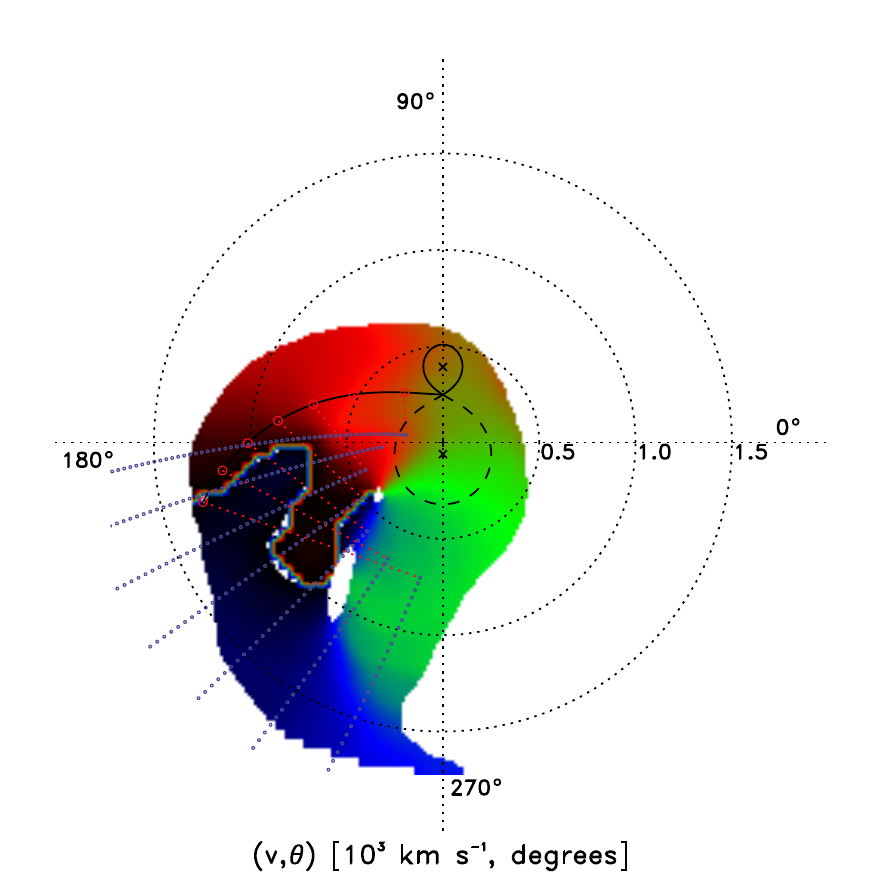} 
 \hspace{-0.250cm}
\includegraphics[height= 7cm, width=7cm]{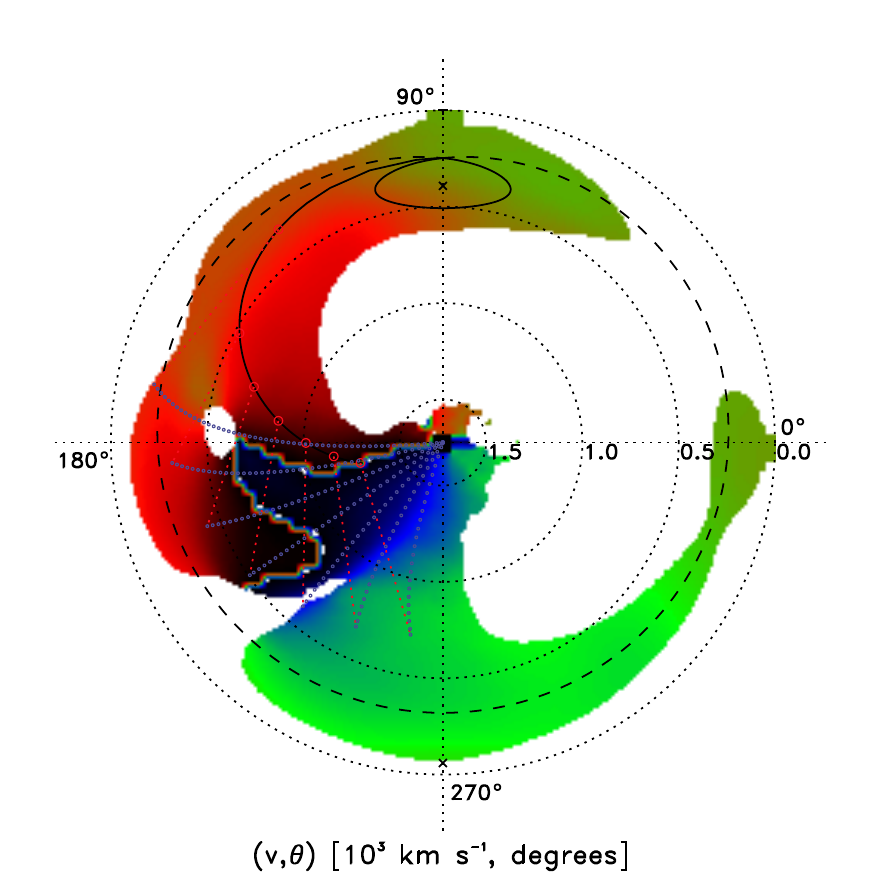}  
\hspace{-0.150cm}
\raisebox{-0.10cm}{\includegraphics[height= 7cm ]{Phase_colorbar_final.png}}  \\
\end{array}$
\end{center}
\vspace{-0.15cm}
\begin{center}$
\begin{array}{cccc}
\includegraphics[width=0.26\textwidth, height= 7cm]{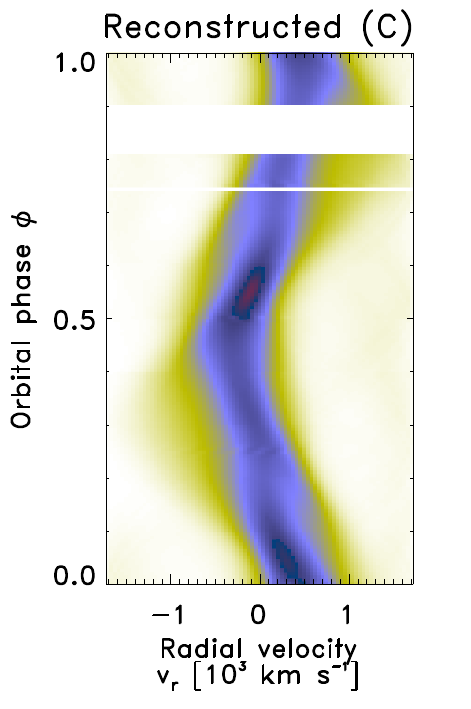} 
 \hspace{-.10cm}
\includegraphics[width=0.26\textwidth, height= 7cm]{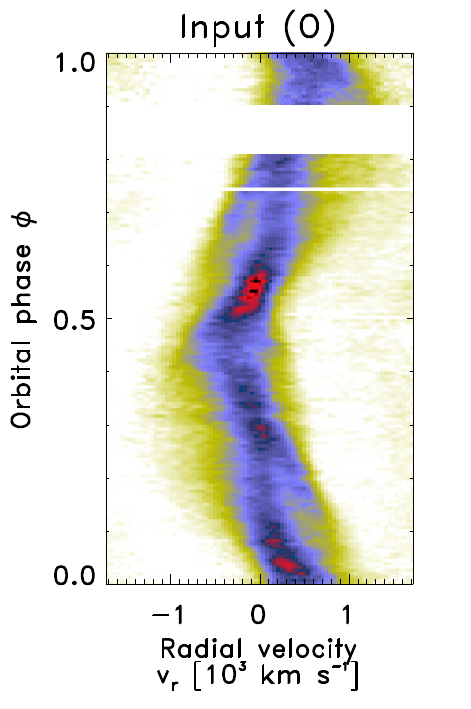} 
 \hspace{-.10cm}
\includegraphics[width=0.26\textwidth, height= 7cm]{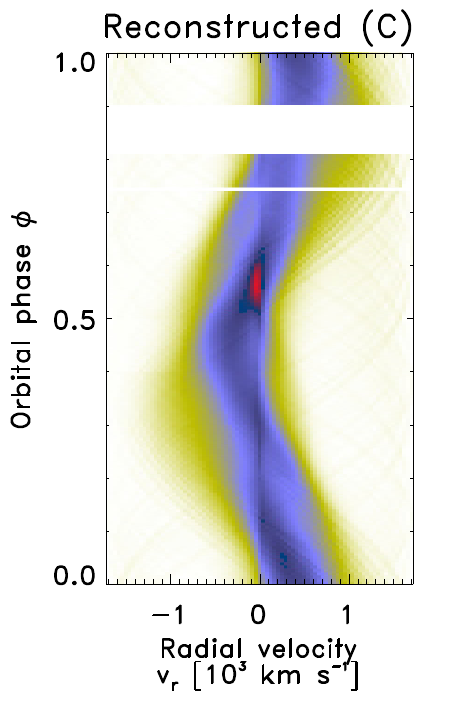} 
 \hspace{-0.15cm}
\raisebox{0.6cm}{\includegraphics[ height= 6.2cm]{Spectra_colorbar_final.png}} \\ 
\end{array}$
\end{center}
\hspace{-0.15cm}
\caption{Standard and inside-out Doppler maps together with the observed and reconstructed trailed spectra based on He~\textsc{ii} 4686 \AA{} emission line from SALT observations. Top row: the standard and inside-out flux modulation amplitude flux Doppler maps. Middle row: the standard and inside-out phase of maximum flux Doppler maps. The phase of maximum flux map shows only pixels where the corresponding modulation amplitude is at least 10\% of the maximum amplitude. Bottom row: the input trailed spectra (centre) with the summed reconstructed trailed spectra for the ten consecutive half-phases for standard (left) and inside-out (right), respectively. }
\label{figure:HeII4686_salt_mod}
\end{figure*}

In this section, we present Doppler maps based on the flux modulation mapping technique described in \cite{2016A&A...595A..47K}, which exploits the principles introduced by \cite{2003MNRAS.344..448S} and \cite{2004MNRAS.348..316P}. 
As already mentioned in \cite{2016A&A...595A..47K}, the modulation mapping technique produces Doppler maps that represent the average, amplitude and phase of the modulated emission. 
Comparison of the reconstructed trailed spectra shown in the bottom row of Figures~\ref{figure:Halpha_mod} and \ref{figure:HeII4686_salt_mod} with the earlier reconstructed trailed spectra shown in Figures~\ref{figure:DopplerHalpha} and \ref{figure:DopplerHeII4686_New} show a significant improvement in reproducing the various components seen in the observed trailed spectra, both in terms of the general shapes of the components and the brightness distribution. 
This, in turn, increases confidence in the reliability of the tomograms.
  
The top row of Figures \ref{figure:Halpha_mod} and \ref{figure:HeII4686_salt_mod} show the standard (left) and inside-out (right) flux modulation amplitude maps of the H$\alpha$ and He~\textsc{ii} 4686 \AA{} emission lines. These maps highlight regions where the emission is modulated over the orbital period due to a combination of projection or occultation effects.
For the H$\alpha$ line, both the standard and inside-out tomograms show flux modulation from the irradiated face of the secondary star, the ballistic and magnetically confined accretion streams. 
Notably, the standard modulation tomogram shows that the irradiated face of the secondary star is the most flux-modulated component. Conversely, the inside-out modulation amplitude map shows that the higher velocity part of the magnetically confined accretion stream is the most flux-modulated component. 
According to \cite{2016A&A...595A..47K}, the secondary dominates the brightness scale in the standard tomogram due to its compacted projection. 
In contrast, its extended projection in the inside-out tomogram makes other emission features appear brighter. 
We note that the other Balmer lines, such as the H$\beta$ line (Figures not shown), show strong flux modulation from the ballistic stream in both the standard and inside-out modulation amplitude maps, instead of the donor star.  

On the other hand, the He~\textsc{ii} 4686 \AA{} tomograms show flux modulation in both the ballistic and magnetically confined accretion streams, with the magnetically confined accretion stream being the most flux-modulated component in both projections (Figure~\ref{figure:HeII4686_salt_mod}). In the standard tomogram, the low-velocity parts of the magnetic stream appear most flux-modulated, whereas in the inside-out tomogram, the high-velocity parts appear most flux-modulated. This, again, is due to the reversed effects between the two projection methods used. 
Compared to the H$\alpha$ tomograms (Figure \ref{figure:Halpha_mod}), both the standard and inside-out tomograms of He~\textsc{ii} 4686 \AA{} show little evidence of flux modulation from the irradiated face of the secondary star. 
In addition, we note strong evidence for the splitting of the magnetically confined accretion stream into two components in the lower left region of the tomograms. These two components are clearly pronounced in both the standard and inside-out flux modulation amplitude tomograms. 
We interpret this as evidence of two distinct magnetic accretion flows, where the magnetic stream splits to feed two accreting poles during the high state, forming two curtain-like structures. Notably, we obtained similar results from observations taken a week later with the SAAO 1.9-m telescope, indicating that these features are not transient events specific to a single dataset, but rather persistent structures observed on two separate occasions with two independent instruments.


\subsubsection{Phase of maximum flux maps}\label{section:phase_of_max_flux}

The middle row of Figures \ref{figure:Halpha_mod} and \ref{figure:HeII4686_salt_mod} show the standard (left) and inside-out (right) phases of maximum flux maps for the H$\alpha$ and He~\textsc{ii}  4686 \AA{} emission lines, respectively. 
These maps indicate the orbital phase at which the various emission components appear the brightest to an observer and have been colour-coded according to phase as follows: 0.0 -- black, 0.25 -- red, 0.5 -- green and 0.75 -- blue, respectively. 
For all lines and, in both the standard and inside-out phases of maximum flux maps, the donor star appears brightest at orbital phase 0.5 (green). 
This is expected since at this phase the irradiated face of the secondary star would be pointing towards the observer. 
The ballistic stream appears brightest to the observer at phase 0.25 (red) for both emission lines. 
This is due to the fact that the observer will have a full view of the inner side of the ballistic stream, that is most exposed to the accreting WD, at this orbital phase. 
Most strikingly—particularly for the He~\textsc{ii} 4686 \AA{} line—is the clear separation of the ballistic stream (red) and the two magnetic streams (blue/black and green), further supporting the distinct detection of two magnetic accretion flows mentioned in Section \ref{section:amplitude_modulation}.

\section{Discussion and Conclusion}\label{section:discussion}

We present phase-resolved optical spectroscopic observations of EF Eri obtained during a high state: the magnitude of EF Eri had risen to $\approx$14.5 mag from the average magnitude of $\leq$16.5 mag reported in the past three decades (i.e. since 1992). 
The average spectrum of EF Eri shows strong emission from the Balmer and He~\textsc{ii} 4686 \AA{} lines consistent with EF Eri being in a high state. 
This is typical of polars in their high state of accretion when mass transfer from the donor star has significantly increased. 
Similar high-state spectra of EF Eri have been presented by different authors in the late 1970s and early 1980s \citep[see e.g.][]{1979ApJ...232L..27G,1979Natur.281...48W,1980A&A....86L..10V}. 

We present the first detailed Doppler tomographic analysis of EF Eri based on our new high-state observations. 
Note that low-state Doppler maps of this system will be presented in due course by \textcolor{blue}{Schwope \& et al. (in prep.)}. 
Generally, the Doppler maps of EF Eri reveal that the ballistic stream dominates the overall emission in the tomograms of the H$\alpha$ line, whereas the magnetically confined accretion stream dominates the overall emission in the tomograms of the He~\textsc{ii} 4686 \AA{} line. 
Several high-state Doppler maps of AM Her systems have been presented in the literature, e.g. HU Aqr \citep{1997A&A...319..894S,2016A&A...595A..47K}, V834 Cen \citep{2004MNRAS.348..316P}, QQ Vul \citep{2000MNRAS.313..533S, 2004ASPC..315..251S} and UZ For \citep{2020MNRAS.492.4298K}. 
The obvious difference between the Doppler maps of EF Eri  when compared with tomograms of other polars in their high state (e.g. HU Aqr \citep{1997A&A...319..894S} and AM Her \citep{2004ASPC..315..251S}) is the absence of strong emission from the irradiated face of the secondary star. 
The Doppler maps of EF Eri show some similarities to that of the He~\textsc{ii} 4686 \AA{} line for V834 Cen \citep{2004MNRAS.348..316P, 2016A&A...595A..47K} and QQ Vul \citep{2000MNRAS.313..533S, 2004ASPC..315..251S} in the sense that they both show strong emission from the ballistic stream and the magnetically confined accretion streams. 
Conversely, the high state Doppler maps of EF Eri are different from those of HU Aqr \citep{1997A&A...319..894S} and AM Her \citep{2004ASPC..315..251S}. 
In particular, the high-state Doppler maps of HU Aqr and AM Her show tomograms dominated by the emission from the secondary star and the ballistic stream with low-level emission from magnetically confined accretion streams. 
Similar Doppler maps of polars showing dominant emission from the secondary star include AR UMa \citep{1999ApJ...525..407S}. 

We also present flux modulation amplitude maps of this system and these show modulation from ballistic and magnetically confined accretion streams for both lines. 
The modulation of the ballistic and magnetically confined streams is expected since EF Eri is a non-eclipsing and mid-inclination system, like V834 Cen \citep{2016A&A...595A..47K}. 
Therefore, the emission from these components will modulate with the orbital phase as these components move in and out of view.  
However, the H$\alpha$ line shows flux modulation from the irradiated face of the secondary star in addition to the ballistic and magnetically confined accretion streams. 
This, again, Even though EF Eri is a non-eclipsing system the flux from the secondary star is expected to modulate with the orbital phase as the irradiated face of the donor star rotate in and out of view.
The He~\textsc{ii} 4686 \AA{} flux modulation amplitude maps reveal that the magnetically confined accretion stream is split into two distinct components as is evident in both the standard and inside-out tomograms. 
These two components are also clearly seen in the phases of maximum flux maps for both projections. 
The first of these components (red and black) follows the ballistic stream closely and traces the motion of the magnetically confined accretion stream as it leaves the orbital plane. 
\cite{2016A&A...595A..47K} reported a similar feature when presenting their flux modulation maps results for V834 Cen. 
For EF Eri, however, a second component (green and blue) which does not follow the ballistic stream is observed in both the standard and inside-out projections. 
This feature has not been reported in the Doppler maps of any other polar and its structure is consistent with the second component of the magnetically confined accretion stream typically feeding the second magnetic pole.   
These results make EF Eri the first polar, to our knowledge, to show rare evidence of two distinct magnetic streams, and an interesting system within the mCV population. 
Further observations are encouraged to study the evolution of the ballistic and magnetic streams through different states.

\section*{Acknowledgements}

The spectroscopic observations reported in this paper were obtained with the Southern African Large Telescope (SALT), under programs 2021-2-LSP-001 (PI: David Buckley) and the 1.9-m telescope of the SAAO in Sutherland. 
The financial assistance of the South African Radio Astronomy Observatory (SARAO) towards this research is hereby acknowledged (\url{www.sarao.ac.za}). 
SBP and DAHB acknowledge research support from the National Research Foundation.

\section*{Data Availability}
 
All data analysed in this work can be made available upon reasonable request to the authors.


\bibliographystyle{mnras}
\bibliography{References} 

\bsp	
\label{lastpage}
\end{document}